\documentclass[twocolumn]{aastex631}

\usepackage{amsmath}
\usepackage{orcidlink}

\usepackage{xcolor}  
\usepackage{etoolbox}

\makeatletter
\patchcmd{\acknowledgments}
  {\begin{internallinenumbers}}
  {\ifnumlines\begin{internallinenumbers}\fi}{}{}
\patchcmd{\endacknowledgments}
  {\end{internallinenumbers}}
  {\ifnumlines\end{internallinenumbers}\fi}{}{}
\patchcmd{\acknowledgements}
  {\begin{internallinenumbers}}
  {\ifnumlines\begin{internallinenumbers}\fi}{}{}
\patchcmd{\endacknowledgements}
  {\end{internallinenumbers}}
  {\ifnumlines\end{internallinenumbers}\fi}{}{}
\makeatother

\begin{document}

\title{The Interpolation Constraint in the RV Analysis of M Dwarfs Using Empirical Templates}

\correspondingauthor{Dhvani Doshi}
\email{dhvani.doshi@mail.mcgill.ca}

\author[0000-0003-3610-3434]{Dhvani Doshi}
\affiliation{Department of Physics, McGill University, 3600 Rue University, Montréal, QC H3A 2T8, Canada}
\affiliation{Trottier Space Institute, McGill University, 3550 Rue University, Montr\'eal, QC H3A 2A7, Canada}

\author[0000-0001-6129-5699]{Nicolas B. Cowan}
\affiliation{Department of Physics, McGill University, 3600 Rue University, Montréal, QC H3A 2T8, Canada}
\affiliation{Trottier Space Institute, McGill University, 3550 Rue University, Montr\'eal, QC H3A 2A7, Canada}
\affiliation{Department of Earth and Planetary Sciences, McGill University, 3450 Rue University, Montréal, QC H3A 0E8, Canada}

\author[0000-0003-3506-5667]{Étienne Artigau}
\affiliation{Institut Trottier de Recherche sur les Exoplanètes, Université de Montréal, 1375 Avenue Thérèse-Lavoie-Roux, Montréal, QC H2V 0B3, Canada}
\affiliation{Observatoire du Mont-Mégantic, Université de Montréal,
Montréal H3C 3J7, Canada}
\affiliation{Département de Physique, Université de Montréal, Montréal QC, Canada}

\author[0000-0001-5485-4675]{René Doyon}
\affiliation{Institut Trottier de Recherche sur les Exoplanètes, Université de Montréal, 1375 Avenue Thérèse-Lavoie-Roux, Montréal, QC H2V 0B3, Canada}
\affiliation{Observatoire du Mont-Mégantic, Université de Montréal,
Montréal H3C 3J7, Canada}
\affiliation{Département de Physique, Université de Montréal, Montréal QC, Canada}

\author[0000-0003-4920-738X]{André M. Silva}
\affiliation{Instituto de Astrofísica e Ciências do Espaço, CAUP, Universidade do Porto, Rua das Estrelas, 4150-762 Porto, Portugal}
\affiliation{Departamento de Física e Astronomia, Faculdade de Ciências, Universidade do Porto, Rua do Campo Alegre, 4169-007 Porto, Portugal}

\author[0000-0002-3212-5778]{Khaled Al Moulla}
\affiliation{Instituto de Astrofísica e Ciências do Espaço, CAUP, Universidade do Porto, Rua das Estrelas, 4150-762 Porto, Portugal}

\author[0000-0002-8669-5733]{Yashar Hezaveh}
\affiliation{Département de Physique, Université de Montréal, Montréal QC, Canada}
\affiliation{Mila—Quebec Artificial Intelligence Institute, Montréal, QC, Canada}

\begin{abstract}
Precise radial velocity (pRV) measurements of M dwarfs in the near-infrared (NIR) rely on empirical templates due to the lack of accurate stellar spectral models in this regime. Templates are assumed to approximate the true spectrum when constructed from many observations or in the high signal-to-noise limit. We develop a numerical simulation that generates SPIRou-like pRV observations from PHOENIX spectra, constructs empirical templates, and estimates radial velocities. This simulation solely considers photon noise and evaluates when empirical templates remain reliable for pRV analysis. Our results reveal a previously unrecognized noise source in templates created from stacking registered observations, establishing a noise floor for such template-based pRV measurements. We find that these templates inherently include distortions in stellar line shapes due to imperfect interpolation at the detector’s sampling resolution. The magnitude of this interpolation error depends on sampling resolution and RV content. Consequently, for stars with higher RV content, such as cooler M-dwarfs, interpolation noise has a larger relative impact, making their performance comparable to hotter M-dwarfs when using detectors with low sampling. For a typical M4V star, SPIRou’s spectral and sampling resolution imposes an RV uncertainty floor of 0.5-0.8 m/s, independent of the star’s magnitude or the telescope’s aperture. These findings reveal a limitation of template-based pRV methods, underscoring the need for improved spectral modeling and better-than-Nyquist detector sampling to reach the next level of RV precision.
\end{abstract}

\section{Introduction} \label{sec:intro}
The radial velocity (RV) technique is a key method for detecting exoplanets by measuring stellar velocity shifts caused by orbiting planets~\citep{wright_radial_2018}. These shifts allow one to estimate a planet's mass, and when considered with a radius estimate from transits, its bulk density, aiding in the identification of habitable Earth-like planets.

M dwarfs are prime targets for exoplanet detection due to their stronger RV signals for small planets~\citep{charbondemin_mdwarfopp}. However, detecting rocky exoplanets is challenging because their RV signals are often just sub-meters per second, requiring extremely precise measurements. Current spectrographs reach precision limits around 10 cm/s to a few m/s, with even the smallest level of noise obscuring planetary signals~\citep{fischer_state_2016,crass_extreme_2021}.

Accurate Doppler shift measurement demands a precise estimate of the intrinsic, non-shifted stellar spectrum, independent of measurement noise. Any source of noise, whether from instrumental effects, telluric contamination, or stellar activity, can introduce spurious signals that obscure or mimic planetary-induced radial velocity variations, making it difficult to detect low-mass planets.

One common method to address RV measurement challenges is the cross-correlation function (CCF) technique, which uses a predefined spectral line list based on the star’s properties~\citep{baranne_coravel_1979,mayor_jupiter-mass_1995,baranne_elodie_1996}. The observed spectrum is cross-correlated with this list to find the velocity shift that best aligns them. Weighting masks further improve precision by giving more influence to lines with stronger RV content~\citep{pepe_coralie_2002}.

While widely used, the CCF method is less effective for M dwarfs~\citep{rainer_stellar_2020,lafarga_carmenes_2020}. Their cool temperatures produce dense, blended spectral features that are difficult to match with line lists. Moreover, M dwarfs emit primarily in the NIR, where RV precision is higher, but current molecular line lists often miss observed lines in this regime, limiting CCF accuracy~\citep{artigau_optical_2018,jahandar_mdwarf}.

To address the limitations of CCF for M-dwarf pRV, data-driven methods construct empirical templates by co-adding multiple observations of the same star. These templates capture the full RV content of the spectrum and have been adopted in pRV measurement algorithms like HARPS-TERRA~\citep{anglada-escude_harps-terra_2012}, NAIRA~\citep{astudillo-defru_harps_2017-1}, SERVAL~\citep{Zechmeister2018}, S-BART~\citep{silva_novel_2022}, and APERO~\citep{cook_apero_2022}. Even line-by-line techniques, such as the LBL algorithm, rely on template matching~\citep{dumusque_measuring_2018,cretignier_measuring_2020,artigau_line-by-line_2022}.

The effectiveness of empirical templates depends on the assumption that they approximate the intrinsic stellar spectrum, particularly when built from high-S/N data or a large number of observations. To test this, we develop a simulation that generates synthetic RV observations, builds templates, and compares template-based RV estimates to those derived from the known intrinsic spectrum, thus quantifying the accuracy and limitations of this method for M dwarfs. We provide the methodology of the simulation in Section~\ref{sec:method}, present the results in Section~\ref{sec:results}, and discuss our findings in Section~\ref{sec:conc}.

\section{Methodology} \label{sec:method}
\begin{figure}
    \centering
    \includegraphics[width=\linewidth]{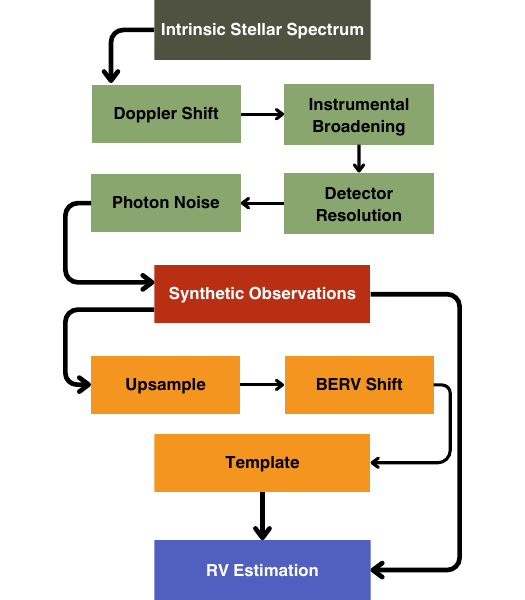}
    \caption{This flowchart describes the steps taken in the numerical simulation to create synthetic RV observations and the template that will then be used for RV estimation.}
    \label{fig:flowchart}
\end{figure}

This section outlines the numerical simulation used to generate radial velocity observations, construct the template, and determine the radial velocities of the observations using either a template or stellar model. These steps are shown in Figure~\ref{fig:flowchart}. The synthetic radial velocity observations are meant to mimic observations from  SpectroPolarimètre InfraRouge (SPIRou) at the Canada-France-Hawaii Telescope on Maunakea. SPIRou has a wavelength range of $0.967\text{-}2.493\,\mathrm{microns}$ microns with a spectral resolution of $70\,000$ and pixel size of $2.28\,\mathrm{km/s}$~\citep{artigau-2014,donati_spirou_2020}.

\subsection{Intrinsic Spectrum Generation}
To generate simulated observations, we first define the intrinsic stellar spectrum, which serves as the ground truth that the template is meant to approximate. We use high-resolution PHOENIX spectra as our intrinsic stellar spectra~\citep{Phoenix}. For the near-infrared range, up to $2500\,\mathrm{nm}$, the PHOENIX models have a sampling resolution of $R_{\mathrm{native}} \approx 500\,000$. This sampling reflects the direct output of the PHOENIX code. 

We simulate M-dwarf spectra with effective temperatures between $2300\,\mathrm{K}$ and $3500\,\mathrm{K}$, extracting the wavelength range corresponding to a given SPIRou order. The simulation is run on a per-order basis, with the selected order provided as an input to define the wavelength range for generating the PHOENIX spectrum. The native PHOENIX spectrum is resampled to a constant sampling resolution of $R=500\,000$, convolved with a Gaussian kernel of width $\sigma_{\mathrm{kernel}} = \frac{1}{2\kappa_{\mathrm{int}}\sqrt{2\ln{2}}}$ (with the spectral resolution set to $\kappa_{\mathrm{int}}=200\,000$), and then interpolated to a finer grid such that the sampling resolution is $R_{\mathrm{int}} = 1\,000\,000$. We use cubic spline interpolation via \texttt{scipy.interpolate.InterpolatedUnivariateSpline} from the SciPy library~\citep{2020SciPy-NMeth}. This function is consistently used for interpolation throughout the numerical simulation.

We define S/N as the signal-to-noise per $1\,\mathrm{km/s}$ domain to enable consistent photon counts across detectors with different sampling resolutions. Since higher-resolution detectors collect fewer photons per pixel, this definition standardizes comparisons. To apply the correct flux density, we normalize the spectrum by its median and scale it by the square of the desired S/N, yielding equivalent results to post-Doppler-shift scaling but simplifying analysis.

This procedure yields a high-resolution synthetic spectrum for RV analysis, assuming a temporally static intrinsic stellar spectrum without activity-induced variability (e.g., flares, spots, or plages). Although this assumption is not physically realistic, the simulation is designed to isolate the effects of numerical methods.

\subsection{Simulated Observations}
To simulate observations, we generate an RV time series by defining the observation window and number of observations, $N_{\mathrm{obs}}$. Observation times mimic realistic cadences, with irregular sampling and gaps. This is achieved by first selecting observation windows at random within the full time span, then randomly sampling observation times within each window. For each time, we compute the Barycentric Earth Radial Velocity (BERV) using the \texttt{helcorr} routine from the \texttt{PyAstronomy.pyasl} module~\citep{pyastronomy}. This is added to a systemic velocity and a circular planetary signal with semi-amplitude $K$.

The intrinsic spectrum is Doppler-shifted according to the predetermined RV curve, generating $N_{\mathrm{obs}}$ shifted spectra. For each RV value, the wavelength shift is calculated using the relativistic Doppler equation. The shifted spectra are then obtained by interpolating the intrinsic spectrum onto the Doppler-shifted wavelength grid.

We simulate SPIRou’s spectral resolution by convolving each spectrum with a Gaussian kernel of width $\sigma_{\mathrm{kernel}} = \frac{1}{2\kappa_{\mathrm{obs}}\sqrt{2\ln{2}}}$, using $\kappa_{\mathrm{obs}} = 70\,000$. Then, we interpolate onto the SPIRou wavelength grid for a given order to place it in SPIRou's sampling resolution. Since the flux density has already been matched, interpolation is used rather than binning.

Finally, photon noise is added to the spectra by modeling the flux in each wavelength bin as a Poisson-distributed variable, with the mean of the distribution equal to the flux in that bin. Random samples are drawn from the Poisson distribution to simulate the statistical noise associated with photon counting. Other sources of noise, such as tellurics, dark pixels, and readout noise, are neglected to isolate the impact of numerical methods.

These steps yield SPIRou-like observations of an M dwarf with a single circular-orbit planet, incorporating RV shifts and noise consistent with a specified S/N.

\subsection{Template Construction}
The template is built by stacking $N_{\mathrm{temp}}$ observations, where $N_{\mathrm{temp}}$ denotes a subset of $N_{\mathrm{obs}}$. The remaining observations are reserved for evaluating the performance of the template. To avoid template anchoring effects, we ensure that the evaluation set is entirely separate from the template construction set. Moreover, by keeping the evaluation set fixed across different choices of $N_{\mathrm{temp}}$, we ensure that all templates are assessed on the same data. This allows for a consistent and unbiased comparison between templates built with varying numbers of input observations.

Each observation used to create the template is resampled onto a higher-resolution grid matching the intrinsic spectrum’s wavelength grid to minimize interpolation errors. This approach is consistent with methods used in other pipelines such as APERO~\citep{cook_apero_2022}, SBART~\citep{silva_novel_2022}, and HARPS-TERRA~\citep{anglada-escude_harps-terra_2012}, which also place their templates on a common reference wavelength grid.

Observations are Doppler-shifted back by their BERV values to align them with the stellar rest frame, reducing line shifts to a fraction of their widths. The template is then taken as the median of these BERV-registered spectra and kept in upsampled space to avoid further interpolation artifacts.

This empirical template represents the intrinsic stellar spectrum convolved to the instrument's spectral resolution. With large \( N_{\mathrm{temp}} \) or high S/N, residual noise is assumed to become negligible, and the template should approximate the intrinsic spectrum.

\subsection{Radial Velocity Determination}
Our RV estimation procedure closely follows the classical template matching procedure outlined in~\cite{silva_novel_2022}. The radial velocities are determined for each order independently, following a least-squares minimization given by
\begin{equation}
     \chi^2 = \sum_{i=1}^{N_{\mathrm{pixels}}} \frac{(F_i - S_i(Y, \lambda_i, v))^2}{\sigma_{F_i}^2 + \sigma_{S_i}^2(Y)},
     \label{eq:chi2}
\end{equation}
where $F_{i}$ is the flux associated with the synthetic observation and $\sigma_{F_i}$ is its associated uncertainty, derived solely from Poisson noise. The residuals of the synthetic observations are calculated with respect to a stellar model $S_{i}(Y,\lambda_i,v)$ and its associated uncertainty, $\sigma_{S_i}(Y)$. The stellar model $S_{i}(Y,\lambda_i,v)$ is either represented by the intrinsic spectrum, $Y = I$, or the constructed template, $Y = T$, and can be written as 
\begin{equation}
    S_{i}(Y,\lambda_i,v) = 
    \begin{cases}
        \mathcal{R} \big( \mathcal{B}(\mathcal{D}(Y,\lambda_i,  \mathrm{BERV} + v)) \big) & \text{if } Y = I,\\
        \mathcal{R} \big( \mathcal{D}(Y, \lambda_i, \mathrm{BERV} + v) \big) & \text{if } Y = T,
    \end{cases}
\end{equation}
where $\mathcal{D}$ represents the relativistic Doppler shift function in which the spectrum is shifted by the observation's BERV and some proposed additional RV shift $v$, $\mathcal{B}$ represents the instrumental broadening function, and $\mathcal{R}$ represents the function to place the spectra in the wavelength grid of the instrument. Similarly, the uncertainty of the stellar template, $\sigma_{S_i}(Y)$ is represented by: 
\begin{equation}
    \sigma_{S_i}(Y) = 
    \begin{cases}
        0 & \text{if } Y = I,\\
        \sigma_{F_i}/\sqrt{N_{\mathrm{temp}}} & \text{if } Y = T.
        .
    \end{cases}
\end{equation}
The $\chi^2$ minimization is performed using SciPy's minimize scalar function using the Brent method~\citep{brent1973algorithms}. The final $\chi^2$ value excludes the outermost $5\%$ of the spectrum in each order to avoid convolution and interpolation artifacts, which could bias the RV estimate. The estimated RV and its associated uncertainty per order is calculated using Equation 3 from~\cite{silva_novel_2022}, where we assume the RV interval for SPIRou is $1\,\mathrm{m/s}$. The RV interval defines the spacing between the central RV estimate and adjacent values, which are used to numerically fit a parabola to the retrieved $\chi^2$ function and estimate the RV uncertainty. Finally, the RV for each observation, including all orders, is calculated using Equation 4 from~\cite{silva_novel_2022}, which is a simple weighted mean. 

\section{Results} \label{sec:results}
This section presents key results from the numerical simulation outlined above. We explain how the template is constructed from spectra that contain deviations from the intrinsic stellar spectrum unrelated to Doppler shifts or photon noise. These deviations, which we attribute to interpolation errors, introduce biases in RV estimates.

\subsection{Variations in Stellar Lines}
Planet-induced Doppler shifts cause subtle changes in stellar spectral lines, often smaller than the lines' widths~\citep{wright_radial_2018}. Thus, any distortion of these lines can bias RV estimates. While this is commonly considered in the context of stellar activity, where features like spots and plages can alter the stellar lines and affect RV estimation, we argue that a similar effect arises in the template creation due to imperfect interpolation at the resolution of our instruments.

To build the template, we upsample observations, interpolating between wavelength bins, before Doppler-shifting them by their BERV. Consider synthetic observations derived from an intrinsic spectrum that has been Doppler shifted, convolved, and down-sampled to the instrument's resolution, but without added photon noise. Ideally, upsampling should preserve the convolved spectral features.

\begin{figure*}
    \centering
    \includegraphics[width=1\linewidth]{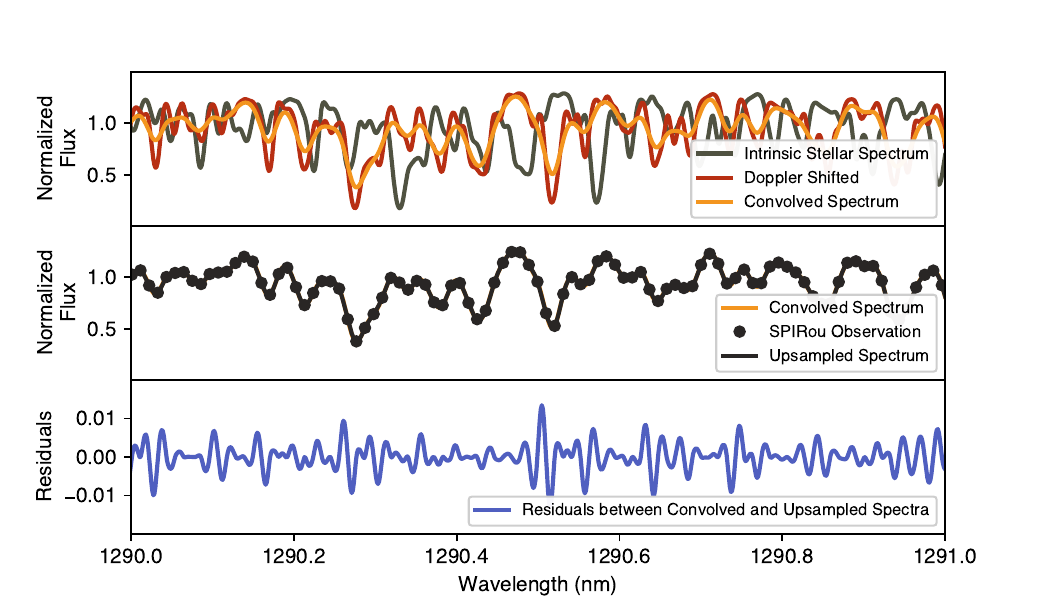}
    \caption{ \textbf{\textit{Top:}} An intrinsic stellar spectrum which is Doppler-shifted and then convolved. \textbf{\textit{Middle:}} The convolved spectrum is down-sampled to SPIRou’s resolution, but no photon noise is added, thus representing a noiseless observation. Here, each black point represents a SPIRou pixel. This synthetic observation is then upsampled using spline interpolation as a precursor to constructing the template.} \textbf{\textit{Bottom:}} Discrepancies between the upsampled and convolved spectrum, indicating that the template will contain altered stellar features.
    \label{fig:interp_error}
\end{figure*}

\begin{figure*}
    \centering
    \includegraphics[width=1\linewidth]{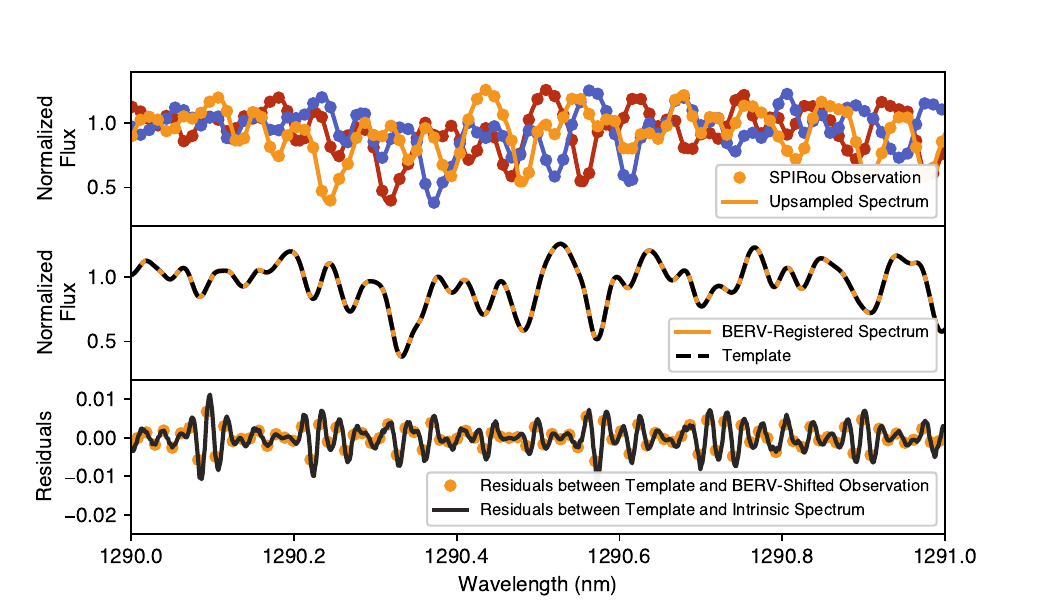}
    \caption{ \textbf{\textit{Top:}} Each color represents a separate noiseless synthetic SPIRou observation along with its corresponding upsampled (spline interpolated) spectrum. For clarity, only three observations are shown. \textbf{\textit{Middle:}} The upsampled spectra are BERV-registered to reduce the line shift to a fraction of their width. The template shown as the dashed black line is the median of 10 BERV-registered spectra. \textbf{\textit{Bottom:}} Discrepancies between the template and instrumentally broadened intrinsic spectrum, demonstrating that the template contains altered stellar features. We further plot the residuals between a BERV-shifted observation and the template interpolated to the same wavelength grid.}
    \label{fig:template_coadd}
\end{figure*}

However, as shown in Figure~\ref{fig:interp_error}, even in this noise-free case, the upsampled spectra show subtle but RV-significant differences from the intrinsic convolved spectrum. Since the upsampled spectra are used to create the template, the templates will retain these deviations from the intrinsic spectrum as seen in Figure~\ref{fig:template_coadd}. In this specific case, the interpolation residuals approach the $\sim1$\% level, even in the absence of photon noise. This places an effective ceiling on the achievable SNR of about $100$, beyond which uncertainties cannot be further reduced. Thus, these residuals can bias RV estimates at high SNR, as they persist independently of photon noise.

This discrepancy arises because interpolation at the instrument’s sampling resolution cannot perfectly reconstruct the original spectrum. A simple test using a Gaussian line profile reveals that cubic spline interpolation tends to underestimate line depth and changes the shape of the line wings. Thus, cubic splines will struggle to reproduce the detailed features present in stellar spectra. 

Importantly, this type of noise would still arise even without directly upsampling the observations, because interpolation is still required after the BERV correction to register the spectra into the same target rest frame to stack them for template construction. More broadly, since the template is constructed from discretely sampled observations but later treated as a continuous model, a certain degree of interpolation is necessary at some stage of the process. As a result, this transition from discrete to continuous representation will introduce subtle variations in spectral line profiles.

\subsection{Significance of Interpolation Error}
\label{sec:zscore_res}
In the previous section, we demonstrated that the template deviates from the intrinsic spectrum even in the absence of measurement noise. Now we will evaluate if this interpolation error is present in the template when estimating RVs. To do this, we calculate the Z-score at each pixel after shifting the stellar model to the retrieved RV. The Z-score measures the residual at each pixel normalized by the combined uncertainty from the observation and the template.

For each order, we compute the standard deviation of these Z-scores, which is equivalent to the square root of the reduced chi-squared defined in Equation~\ref{eq:chi2}. A standard deviation exceeding 1 suggests that the template contains additional noise not accounted for by photon statistics alone.

We test this using an M9V PHOENIX model, simulating spectra across various sampling resolutions, including SPIRou’s native resolution, and varying S/N. Each test uses $10$ noise realizations, templates built from $100$ observations, and is evaluated on $100$ independent observations across all SPIRou orders.

\begin{figure}
    \centering
    \includegraphics[width=0.45\textwidth]{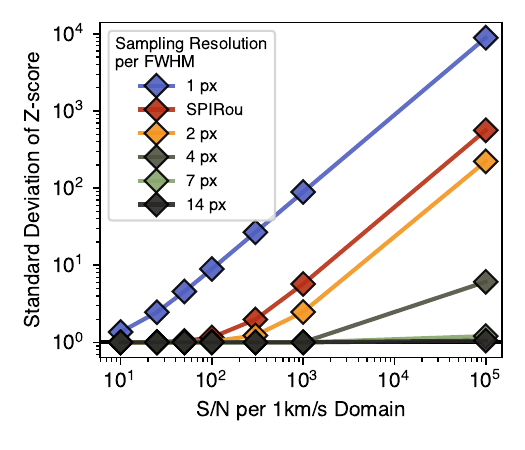}
    \includegraphics[width=0.45\textwidth]{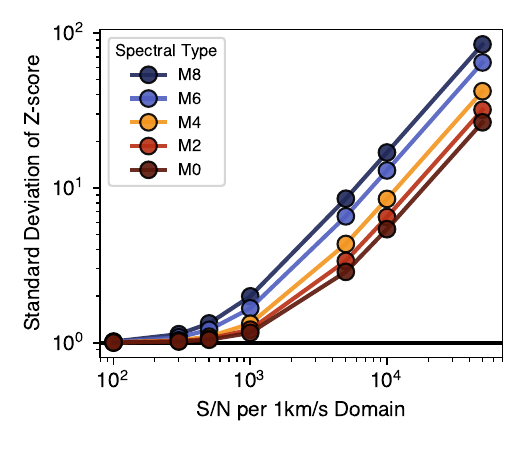}
    \caption{The standard deviation of the Z-scores between the observation and the template is shown for different cases. A standard deviation greater than one indicates interpolation errors. \textbf{\textit{Top:}} The significance of interpolation error depends on both sampling resolution and photon noise, with each line representing a different sampling resolution. \textbf{\textit{Bottom:}} The standard deviation of the Z-scores for different M-dwarf types at SPIRou’s native resolution shows that interpolation errors increase for cooler M dwarfs with higher RV content.} 
    \label{fig:zscore}
\end{figure}

The top panel of Figure~\ref{fig:zscore} shows the mean Z-score standard deviation across all pixels, orders, noise realizations, and observations. For a given sampling resolution, the standard deviation remains near $1$ until the S/N surpasses a certain threshold. This threshold reflects the magnitude of the interpolation error at that sampling resolution. Below this threshold, photon noise dominates the template. However, above this threshold, the standard deviation increases, indicating that interpolation error begins to dominate in this regime.

Higher sampling resolution lowers interpolation error, reflected in decreasing Z-score deviations with increased sampling. These results are independent of $N_{\mathrm{temp}}$ as the Z-score accounts for photon noise in the template. For comparison, using the intrinsic spectrum instead of a template yielded Z-score deviations of about $1$ across all S/Ns and resolutions, as expected.

\subsection{Dependence on Stellar Type}
\label{sec:stelltype}
The performance of interpolation functions should vary depending on the underlying spectrum. The magnitude of the error should decrease if there are fewer spectral lines or more broadened features.

To test this, we repeat the experiment outlined in Section~\ref{sec:zscore_res}, using only SPIRou's native resolution sampling. Here, we use PHOENIX models varying $T_{\mathrm{eff}}$ while holding $\log(g)$ and $[\mathrm{Fe/H}]$ constant. 

In the bottom panel of Figure~\ref{fig:zscore}, we compare the standard deviation of the Z-scores as described in Section~\ref{sec:zscore_res}. As we increase $T_{\mathrm{eff}}$, the number of spectral lines decreases, leading to fewer deep and complex features in the spectrum. Since these features are more susceptible to interpolation errors, their absence reduces the magnitude of the overall interpolation error. This demonstrates that the interpolation error is dependent on the RV content in the intrinsic spectrum. Consequently, while cooler M dwarfs are expected to yield lower RV uncertainties due to their richer RV information, their dense spectral features also amplify interpolation errors, potentially biasing RV estimates.

\subsection{Impact on RV Estimation}
We’ve shown that interpolation errors can dominate over photon noise at high S/N. Here, we evaluate how these errors impact RV estimates under more realistic S/N conditions. To do this, we compare RVs retrieved using either the intrinsic stellar spectrum or the template to the injected RVs. Since the template is constructed from the median of multiple observations, it includes an inherent Doppler shift, so we correct for this by fitting a sine curve to the retrieved RVs and removing the vertical offset.

\begin{figure*}
    \centering
    \includegraphics[width=1\linewidth]{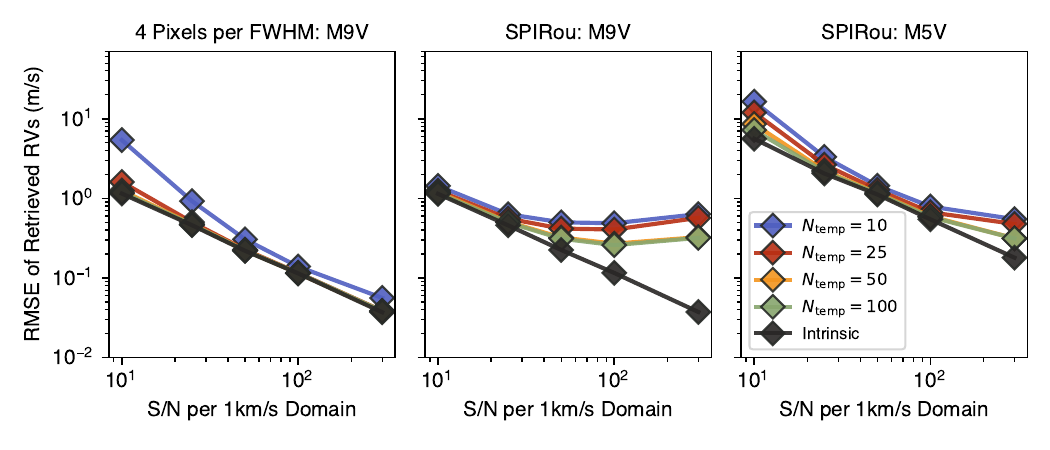}
    \caption{The root mean squared error of the estimated RVs for $100$ evaluation observations for templates created with a varying number of observations, $N_{\mathrm{temp}}$, and with the intrinsic stellar spectrum. In the first panel, we plot the RMSE for a hypothetical instrument with $4$ pixels per FWHM observing an M9 star. We see that the RMSE of the template follows Poisson statistics. In the second panel, we observe the same M9 star but at the sampling resolution of SPIRou. Here we see an RMSE plateau at high S/N due to interpolation error. In the last panel, we compare this to observations of an M5 star with SPIRou.}
    \label{fig:residuals}
\end{figure*}

We simulate observations of an M9V star at various S/N values. We build a template from different $N_{\mathrm{temp}}$ values and evaluate our RV estimates on $100$ independent observations. We run the experiment for $10$ random noise instances. In Figure~\ref{fig:residuals}, we compare the root mean squared error (RMSE) of the RV estimates for the experiment outlined above between a $4$ pixels per FWHM instrument and SPIRou. 

At high sampling resolution, the RMSE follows Poisson scaling as S/N and 
$N_{\mathrm{temp}}$ increase. However, for SPIRou, the RMSE plateaus at high S/N, indicating that interpolation error limits precision. This effect is more pronounced for the M9V star, while the M5V star shows milder signs of interpolation error at the same S/N. The slight upward trend seen in the middle panel of Figure~\ref{fig:residuals} arises because the template biases the RV estimation in the noiseless regime due to interpolation artifacts. Thus, adding noise does not simply degrade the measurement but alters the bias while increasing the variance in a non-trivial way.

To further visualize the impact of interpolation errors on retrieved RVs, we generate synthetic SPIRou observations using an M9V at three S/N points. We plot the retrieved RVs for $100$ evaluation points using a template generated from $50$ observations and the intrinsic spectrum.

In Figure~\ref{fig:RV_curve}, we see that for an S/N of $100\,000$ the RVs found using the template have clear offsets from the true RVs when using SPIRou. Since the photon noise is negligible in this case, the offsets are simply from the interpolation error. These offsets are random and resemble stellar activity because the template inherits spectral line variations from the spectra used to construct it. At an S/N of $300$, the RV offsets are influenced not only by photon noise but also remain significantly affected by interpolation error. Only at very low S/N does photon noise become large enough to drown the interpolation error. This behaviour is not evident for a detector with $4$ pixels per FWHM, as there are no clear offsets in the S/N of $100\,000$ case and thus no systematic offsets in the other S/N scenarios as well. 

\section{Discussion and Conclusion}\label{sec:conc}
Imperfect interpolation at the detector's resolution introduces noise in empirical templates, causing RV offsets that resemble stellar jitter or activity. The magnitude of this noise depends on the interpolation function, sampling resolution, and stellar type. This noise is independent of the presence of a planetary signal and may arise in template-based RV measurements taken from simply stacking registered observations. As such, the resulting offsets are also independent of RV semi-amplitude and consequently have minimal impact on high-amplitude systems but pose challenges for low-amplitude systems, i.e., small planets or long orbits.

In the high S/N regime, these offsets lead to biased RV estimates in synthetic SPIRou observations, as only photon noise is typically considered in templates for RV estimation. To account for this, we approximate the flux noise due to interpolation errors, which can be added in quadrature to photon noise when computing total uncertainty, such that,
\begin{equation}
    \sigma_S(T) = \sqrt{\sigma_{\text{photon}}^2 + \sigma_{\text{interp}}^2},
\end{equation}
where $\sigma_{\text{interp}}$ depends on the sampling resolution, RV content, number of observations used to construct the template, and S/N per pixel. We provide $\sigma_{\text{interp}}$ for SPIRou, assuming a template from $100$ observations.

\begin{figure}
    \centering
    \includegraphics[width=1\linewidth]{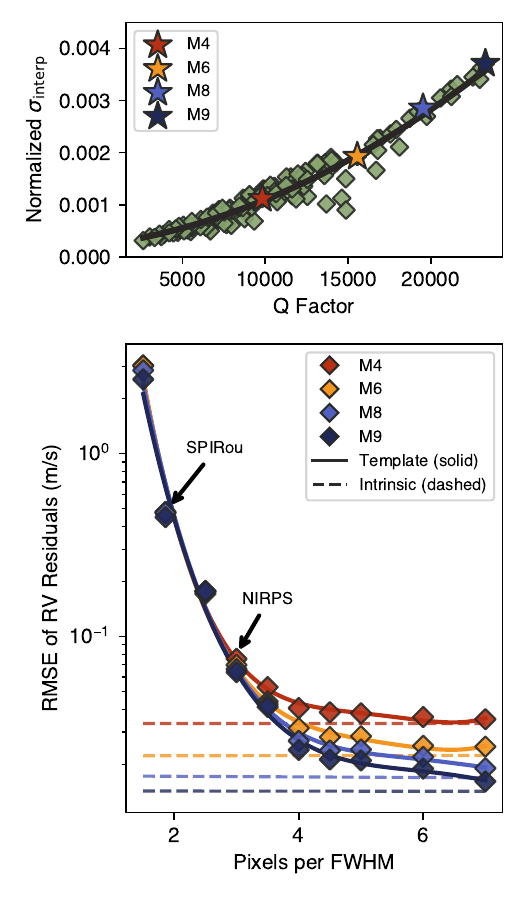}
    \caption{\textbf{\textit{Top:}} The normalized interpolation error, represented by the term in square brackets in Equation~\ref{eq:siginterp}, is plotted against the spectral quality factor and follows a second-order polynomial shown by the black line. Diamonds indicate various spectra, with the positions of example M4, M6, M8, and M9 stars highlighted. \textbf{\textit{Bottom:}} This provides the RV uncertainty floor due to interpolation error for a given spectral type and sampling resolution. RMSE in RV estimates for where S/N is $1000$ per $1\,\mathrm{km/s}$ domain across sampling resolutions and spectral types. Dashed lines show RV precision using the intrinsic spectrum. Template-based estimates are limited to $\sim0.5\,\mathrm{m/s}$ for SPIRou. A 4th-order polynomial fit captures template-based interpolation error.}
    \label{fig:unc}
\end{figure}

To calculate $\sigma_{\text{interp}}$, we generate synthetic observations using a spectrum with a quality factor $Q$, using the formalism from~\cite{bouchy_fundamental_2001}, with no photon noise added. A higher quality factor indicates a spectrum with more lines and greater RV content. The standard deviation of residuals between the template and individual observations is computed for over $500$ observations. The mean of these residuals defines $\sigma_{\text{interp}}$. The relationship between $Q$ and $\sigma_{\text{interp}}$ is shown in the top panel of Figure~\ref{fig:unc}.

The normalized interpolation error is multiplied by the S/N per pixel squared to obtain $\sigma_{\text{interp}}$ for uncertainty calculations. The empirical relationship between $Q$ and $\sigma_{\text{interp}}$ is thus given by:
\begin{equation}
\sigma_{\text{interp}} = (\text{S/N})^2 \cdot \left[ \frac{1}{10^4} \left( 4.31 \left( \frac{Q}{10^4} \right)^2 + 4.39 \frac{Q}{10^4} + 2.23 \right) \right]
\label{eq:siginterp}
\end{equation}
Incorporating $\sigma_{\text{interp}}$ into RV analyses does not address the root cause of interpolation error but helps mitigate bias in the high-S/N regime by effectively attenuating the weights of high-S/N pixels. Using this uncertainty estimation can be advantageous due to its dependence on the spectrum’s quality factor, which makes the result independent of how well PHOENIX models match real M-dwarf spectra.

Alternatively, we also provide an overall RV bias from this interpolation error, which is parameterized by spectral type, sampling resolution, and number of template observations. We generate synthetic spectra at an S/N of $1000$ per $1\mathrm{km/s}$ domain for various M-dwarf types and estimate the RVs using $50$-observation templates and compute the RMSE of RV offsets due solely to interpolation error for over $500$ different observations.

As shown in Figure~\ref{fig:unc}, the RMSE decreases with higher sampling resolution, consistent with the $\mathcal{O}(N^{-4})$ scaling of cubic spline interpolation~\citep{burden2010numerical}. Cooler M dwarfs with a higher Q exhibit greater interpolation errors due to denser spectral features. At low resolutions, this offsets their intrinsic RV advantage, while at high resolutions, the expected precision trend from~\cite{bouchy_fundamental_2001} re-emerges. It should be noted that spectra with a lower Q are still affected by interpolation errors, they simply require higher S/N to reveal the effect.

Overall, SPIRou’s sampling imposes an RV floor of $\sim0.5\,\mathrm{m/s}$, regardless of stellar type or telescope aperture. In contrast, finer-sampled instruments like NIRPS reduce this to $\sim0.07\,\mathrm{m/s}$, highlighting the need for sampling beyond the Nyquist rate to reach sub-$0.1\,\mathrm{m/s}$ precision.

Moreover, in this simulation, we assume a Gaussian point spread function (PSF), which allows a cubic spline to reasonably approximate the underlying stellar spectrum. However, this approximation doesn’t hold perfectly true in reality and many fibre-fed spectrographs exhibit PSFs that are closer to top-hat profiles. Cubic splines are not well-matched to such shapes, and their ability to recover the true spectral content in these cases is unclear, though likely degraded~\citep{burden2010numerical}. As a result, the interpolation errors reported here may be underestimated for instruments with non-Gaussian PSFs, where spectral modeling is inherently more difficult.

In the NIR, two dominant noise sources limiting pRV precision for M-dwarfs are telluric contamination and stellar activity~\citep{fischer_state_2016}. These not only affect individual observations but also degrade empirical templates, compounding errors beyond interpolation error. It is therefore important to assess interpolation error in the context of these effects. Telluric residuals have been shown to induce RV scatter of several m/s in CRIRES data~\citep{bean_crires_2010}, and simulations suggest systematic errors of $1\,\mathrm{m/s}$ for Sun-like stars and up to $2\,\mathrm{m/s}$ for M-dwarfs~\citep{wang_characterizing_2022, latouf_characterizing_2022}. Stellar jitter, as observed in a CARMENES survey of $98$ M-dwarfs, can produce residuals of $3\,\mathrm{m/s}$~\citep{lafarga_carmenes_2021}. However, these estimates often neglect modern mitigation strategies, which continue to improve. As such techniques advance, interpolation error may become a dominant or comparable systematic, limiting further gains in RV precision.

For example, the presence of this interpolation noise floor may help explain the RV scatter observed in seemingly inactive stars. For instance,~\cite{artigau_line-by-line_2022} reported a dispersion of $2.59\,\mathrm{m/s}$ for Barnard's star using SPIRou data analyzed with a line-by-line algorithm, despite a predicted photon noise of only $1\,\mathrm{m/s}$. While other sources, such as modal noise, may contribute~\citep{mahadevan_modal}, interpolation error remains a plausible and significant factor in this discrepancy.

Furthermore,~\cite{figueira_radial_2016} emphasized that sampling above the Nyquist rate is essential for effective telluric correction, as undersampling worsens the ability to flag and remove atmospheric features. This underscores the dual role of high sampling resolution in mitigating both telluric and interpolation errors. Thus, efforts to improve RV precision in the NIR must prioritize detector sampling above the Nyquist rate.

Beyond improved detector sampling, these interpolation artifacts highlight the need for better spectral modeling. Interpolation noise is difficult to eliminate entirely in standard template-based methods, since templates are constructed from discretely sampled observations and therefore cannot perfectly capture the underlying continuous stellar spectrum required for high RV precision. This interpolation issue is further highlighted in~\cite{hogg}, which proposes a forward-modeling approach to circumvent the need for interpolation. In this framework, the stellar spectrum is expressed as a sum of basis functions whose parameters are optimized through a likelihood analysis of the data, leaving the observations untouched, conceptually similar to techniques such as \emph{wobble}~\citep{bedell}. For the standard template-matching technique, mitigation strategies can include smoothing spectra before co-adding them to build the template, or sampling from multiple observations across a range of BERV values to better capture the underlying spectrum. These approaches, however, require further testing to establish whether they can effectively reduce interpolation errors. Ultimately, progress in spectral modeling, adoption of forward-modeling techniques, and improved knowledge of the instrument’s PSF will be essential for fully extracting the true RV content and achieving the next level of precision.

In conclusion, we developed a simulation to generate synthetic pRV observations, allowing us to test the performance of the template in various regimes and compare it to pRV analysis using the intrinsic spectrum. Our findings revealed that the template introduced interpolation errors in synthetic SPIRou observations, which caused offsets in the retrieved RVs. 

This highlights a limitation in the precision that can be achieved with certain instruments when using the standard template technique for RV retrievals. These limitations arise because templates created by stacking registered observations rely on discretely sampled observations to approximate a continuous stellar spectrum. In regimes where spectral lines are blended or the underlying features are not well understood, as is common in M dwarfs, multiple continuous spectra can give rise to the same sampled data, making the interpolation inherently ambiguous. Even with perfect knowledge of the instrumental profile, this ambiguity persists due to the information loss from finite sampling.

While we provide general estimates of the interpolation error for a given instrument, we strongly encourage using this simulation to validate whether a template for a specific star at a given S/N is an adequate approximation for accurate RV retrievals.

\begin{acknowledgements}
This work was supported by the Trottier Institute for Research on Exoplanets (IREx) and the Trottier Space Institute (TSI). The authors would like to thank the anonymous referee for their helpful feedback, which greatly improved this work. We would also like to thank Neil Cook for providing information regarding SPIRou. D.D is supported by the Fonds de recherche du Québec – Nature et technologies (FRQNT). NBC acknowledges support from an NSERC Discovery Grant, a Tier 2 Canada Research Chair, and an Arthur B. McDonald Fellowship. AMS was funded by the European Union (ERC, FIERCE, 101052347). Views and opinions expressed are however those of the author(s) only and do not necessarily reflect those of the European Union or the European Research Council. Neither the European Union nor the granting authority can be held responsible for them. AMS was also supported by FCT - Fundação para a Ciência e a Tecnologia through national funds by these grants: UIDB/04434/2020 DOI: 10.54499/UIDB/04434/2020, UIDP/04434/2020 DOI: 10.54499/UIDP/04434/2020, PTDC/FIS-AST/4862/2020, UID/04434/2025. KA acknowledges support from the Swiss National Science Foundation (SNSF) under the Postdoc Mobility grant P500PT\_230225. 
\end{acknowledgements}

\appendix
\renewcommand{\thefigure}{A.\arabic{figure}}
\setcounter{figure}{0}  
\section{Additional Figures}
This appendix contains the plot to demonstrate that the interpolation error resembles stellar jitter.
\begin{figure*}
    \centering
    \includegraphics[trim=10 40 10 50, clip,width=1\linewidth]{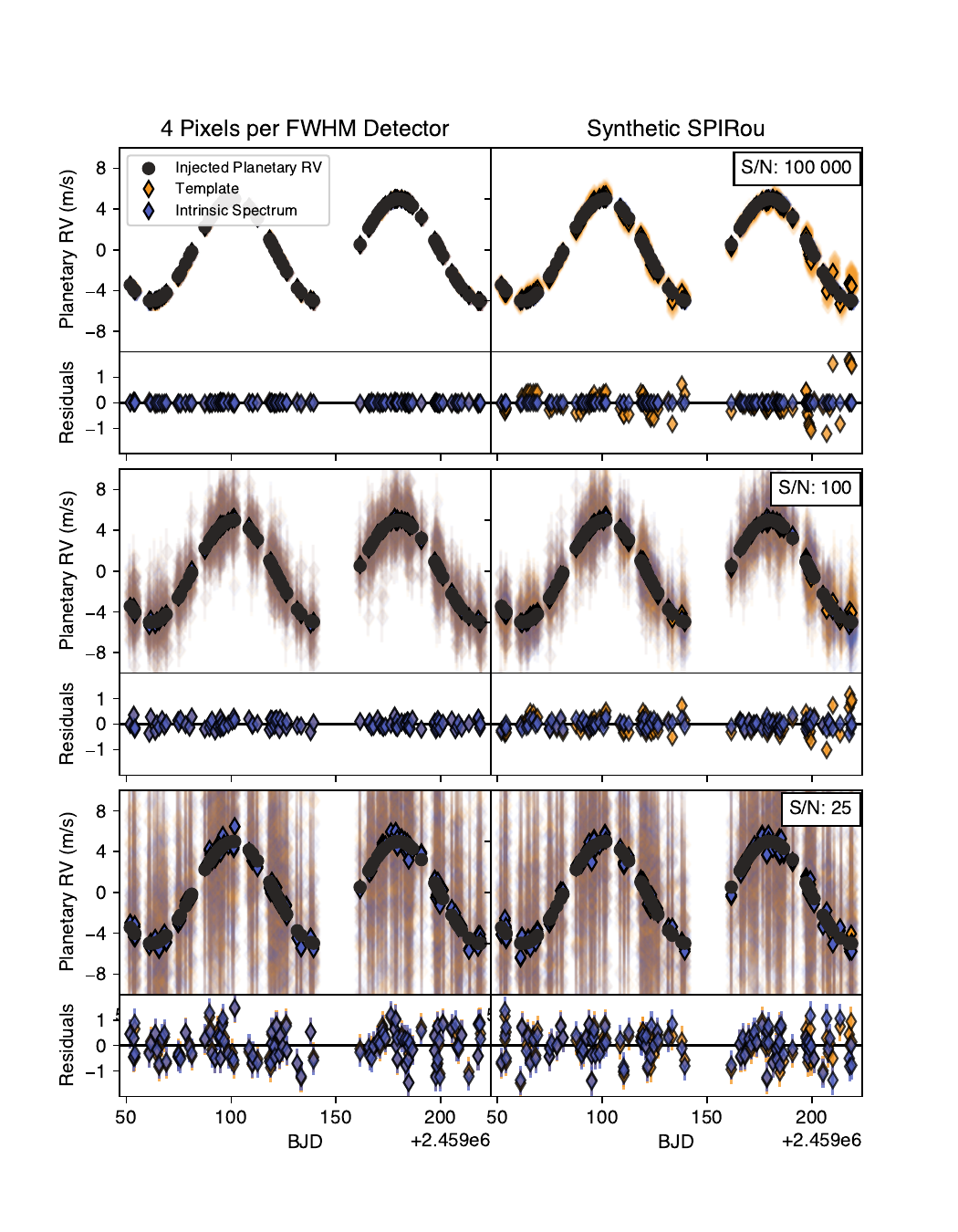}
    \caption{Retrieved planetary RV curves using either the intrinsic stellar spectrum or a template built from $50$ observations, shown for three S/N regimes. There is also a BERV that contributes to the overall RV curve, however, only the planetary RV is plotted. Results are displayed for both SPIRou's native sampling and an instrument with $4$ pixels per FWHM. Semi-transparent points show per-order RVs; opaque points show the order-combined RV. At SPIRou resolution, interpolation error from the template introduces random RV offsets, especially at high and moderate S/Ns, which are absent at finer sampling or when using the intrinsic spectrum.}
    \label{fig:RV_curve}
\end{figure*}

\bibliography{ref}{}
\bibliographystyle{aasjournal}

\end{document}